\documentstyle[twoside]{article}

\catcode`\@=11
\long\def\@makefntext#1{
\protect\noindent \hbox to 3.2pt {\hskip-.9pt  
$^{{\eightrm\@thefnmark}}$\hfil}#1\hfill}		

\def\thefootnote{\fnsymbol{footnote}}
\def\@makefnmark{\hbox to 0pt{$^{\@thefnmark}$\hss}}	
	
\def\ps@myheadings{\let\@mkboth\@gobbletwo
\def\@oddhead{\hbox{}
\rightmark\hfil\eightrm\thepage}   
\def\@oddfoot{}\def\@evenhead{\eightrm\thepage\hfil
\leftmark\hbox{}}\def\@evenfoot{}
\def\sectionmark##1{}\def\subsectionmark##1{}}

\oddsidemargin=\evensidemargin
\renewcommand{\thefootnote}{\fnsymbol{footnote}}

\newcounter{sectionc}\newcounter{subsectionc}\newcounter{subsubsectionc}
\renewcommand{\section}[1] {\vspace{12pt}\addtocounter{sectionc}{1} 
\setcounter{subsectionc}{0}\setcounter{subsubsectionc}{0}\noindent 
	{\tenbf\thesectionc. #1}\par\vspace{5pt}}
\renewcommand{\subsection}[1] {\vspace{12pt}\addtocounter{subsectionc}{1} 
	\setcounter{subsubsectionc}{0}\noindent 
	{\bf\thesectionc.\thesubsectionc. {\kern1pt \bfit #1}}\par\vspace{5pt}}
\renewcommand{\subsubsection}[1] {\vspace{12pt}\addtocounter{subsubsectionc}{1}
	\noindent{\tenrm\thesectionc.\thesubsectionc.\thesubsubsectionc.
	{\kern1pt \tenit #1}}\par\vspace{5pt}}
\newcommand{\nonumsection}[1] {\vspace{12pt}\noindent{\tenbf #1}
	\par\vspace{5pt}}

\newcounter{appendixc}
\newcounter{subappendixc}[appendixc]
\newcounter{subsubappendixc}[subappendixc]
\renewcommand{\thesubappendixc}{\Alph{appendixc}.\arabic{subappendixc}}
\renewcommand{\thesubsubappendixc}
	{\Alph{appendixc}.\arabic{subappendixc}.\arabic{subsubappendixc}}

\renewcommand{\appendix}[1] {\vspace{12pt}
        \refstepcounter{appendixc}
        \setcounter{figure}{0}
        \setcounter{table}{0}
        \setcounter{lemma}{0}
        \setcounter{theorem}{0}
        \setcounter{corollary}{0}
        \setcounter{definition}{0}
        \setcounter{equation}{0}
        \renewcommand{\thefigure}{\Alph{appendixc}.\arabic{figure}}
        \renewcommand{\thetable}{\Alph{appendixc}.\arabic{table}}
        \renewcommand{\theappendixc}{\Alph{appendixc}}
        \renewcommand{\thelemma}{\Alph{appendixc}.\arabic{lemma}}
        \renewcommand{\thetheorem}{\Alph{appendixc}.\arabic{theorem}}
        \renewcommand{\thedefinition}{\Alph{appendixc}.\arabic{definition}}
        \renewcommand{\thecorollary}{\Alph{appendixc}.\arabic{corollary}}
        \renewcommand{\theequation}{\Alph{appendixc}.\arabic{equation}}
        \noindent{\tenbf Appendix \theappendixc #1}\par\vspace{5pt}}
\newcommand{\subappendix}[1] {\vspace{12pt}
        \refstepcounter{subappendixc}
        \noindent{\bf Appendix \thesubappendixc. {\kern1pt \bfit #1}}
	\par\vspace{5pt}}
\newcommand{\subsubappendix}[1] {\vspace{12pt}
        \refstepcounter{subsubappendixc}
        \noindent{\rm Appendix \thesubsubappendixc. {\kern1pt \tenit #1}}
	\par\vspace{5pt}}

\newcommand{\textlineskip}{\baselineskip=13pt}
\newcommand{\smalllineskip}{\baselineskip=10pt}

\def\eightcirc{
\begin{picture}(0,0)
\put(4.4,1.8){\circle{6.5}}
\end{picture}}
\def\eightcopyright{\eightcirc\kern2.7pt\hbox{\eightrm c}} 

\newcommand{\copyrightheading}[1]
	{\vspace*{-2.5cm}\smalllineskip{\flushleft
	{\footnotesize International Journal of Modern Physics A, #1}\\
	{\footnotesize $\eightcopyright$\, World Scientific Publishing
	 Company}\\
	 }}

\def\abstracts#1#2#3{{
	\centering{\begin{minipage}{4.5in}\baselineskip=10pt\footnotesize
	\parindent=0pt #1\par 
	\parindent=15pt #2\par
	\parindent=15pt #3
	\end{minipage}}\par}} 

\renewenvironment{thebibliography}[1]
	{\frenchspacing
	 \ninerm\baselineskip=11pt
	 \begin{list}{\arabic{enumi}.}
	{\usecounter{enumi}\setlength{\parsep}{0pt}
	 \setlength{\leftmargin 12.7pt}{\rightmargin 0pt} 
	 \setlength{\itemsep}{0pt} \settowidth
	{\labelwidth}{#1.}\sloppy}}{\end{list}}

\newcounter{itemlistc}
\newcounter{romanlistc}
\newcounter{alphlistc}
\newcounter{arabiclistc}

\newcommand{\fcaption}[1]{
        \refstepcounter{figure}
        \setbox\@tempboxa = \hbox{\footnotesize Fig.~\thefigure. #1}
        \ifdim \wd\@tempboxa > 5in
           {\begin{center}
        \parbox{5in}{\footnotesize\smalllineskip Fig.~\thefigure. #1}
            \end{center}}
        \else
             {\begin{center}
             {\footnotesize Fig.~\thefigure. #1}
              \end{center}}
        \fi}

\newcommand{\tcaption}[1]{
        \refstepcounter{table}
        \setbox\@tempboxa = \hbox{\footnotesize Table~\thetable. #1}
        \ifdim \wd\@tempboxa > 5in
           {\begin{center}
        \parbox{5in}{\footnotesize\smalllineskip Table~\thetable. #1}
            \end{center}}
        \else
             {\begin{center}
             {\footnotesize Table~\thetable. #1}
              \end{center}}
        \fi}

\def\@citex[#1]#2{\if@filesw\immediate\write\@auxout
	{\string\citation{#2}}\fi
\def\@citea{}\@cite{\@for\@citeb:=#2\do
	{\@citea\def\@citea{,}\@ifundefined
	{b@\@citeb}{{\bf ?}\@warning
	{Citation `\@citeb' on page \thepage \space undefined}}
	{\csname b@\@citeb\endcsname}}}{#1}}

\newif\if@cghi
\def\cite{\@cghitrue\@ifnextchar [{\@tempswatrue
	\@citex}{\@tempswafalse\@citex[]}}
\def\citelow{\@cghifalse\@ifnextchar [{\@tempswatrue
	\@citex}{\@tempswafalse\@citex[]}}
\def\@cite#1#2{{$\null^{#1}$\if@tempswa\typeout
	{IJCGA warning: optional citation argument 
	ignored: `#2'} \fi}}

\def\pmb#1{\setbox0=\hbox{#1}
	\kern-.025em\copy0\kern-\wd0
	\kern.05em\copy0\kern-\wd0
	\kern-.025em\raise.0433em\box0}

\def\fnt#1#2{\footnotetext{\kern-.3em
	{$^{\mbox{\scriptsize #1}}$}{#2}}}

\def\fpage#1{\begingroup
\voffset=.3in
\thispagestyle{empty}\begin{table}[b]\centerline{\footnotesize #1}
	\end{table}\endgroup}

\def\runninghead#1#2{\pagestyle{myheadings}
\markboth{{\protect\footnotesize\it{\quad #1}}\hfill}
{\hfill{\protect\footnotesize\it{#2\quad}}}}
\font\tenrm=cmr10
\font\tenit=cmti10 
\font\tenbf=cmbx10
\font\bfit=cmbxti10 at 10pt
\font\ninerm=cmr9

\font\eightrm=cmr8

\def\qed{\hbox{${\vcenter{\vbox{			
   \hrule height 0.4pt\hbox{\vrule width 0.4pt height 6pt
   \kern5pt\vrule width 0.4pt}\hrule height 0.4pt}}}$}}

\renewcommand{\thefootnote}{\fnsymbol{footnote}}	

\newcommand{\half}{\mbox{$\frac{1}{2}$}}
\newcommand{\slaf}{\mbox{$\frac{1}{2\pi}$}}
\newcommand{\OL}[1]{ \hspace{1pt}\overline{\hspace{-1pt}#1
   \hspace{-1pt}}\hspace{1pt} }
\input{epsf}

\begin{document}

\runninghead{${\cal N} = 2$ Gauge/Gravity Duals} {${\cal N} = 2$
Gauge/Gravity Duals}

\normalsize\textlineskip
\thispagestyle{empty}
\setcounter{page}{1}

\copyrightheading{}			

\vspace*{0.88truein}

\fpage{1}
\centerline{\bf ${\cal N}=2$ GAUGE-GRAVITY DUALS}
\vspace*{0.37truein}
\centerline{\footnotesize JOSEPH POLCHINSKI}
\vspace*{0.015truein}
\centerline{\footnotesize\it Institute for Theoretical Physics, University
of California}
\baselineskip=10pt
\centerline{\footnotesize\it Santa Barbara, California 93106-4030,
USA}
\vspace*{0.225truein}

\vspace*{0.21truein}
\abstracts{We study the ${\cal N} = 2$ analog of the Klebanov-Strassler
system.  We first review the resolution of singularities by the enhan\c con
mechanism, and the physics of fractional branes on an orbifold.  We then
describe the exact ${\cal N} = 2$ solution.  This exhibits a duality cascade
as in the ${\cal N} = 1$ case, but the singularity resolution is the
characteristic ${\cal N} = 2$ enhan\c con.  We discuss some related
solutions and open issues.}{}{}


\vspace*{1pt}\textlineskip	
\section{Introduction}	
\vspace*{-0.5pt}
\noindent
Maldacena's duality\cite{Maldacena} between IIB string theory on $AdS_5 \times
S^5$ and
${\cal N} = 4$ supersymmetric Yang-Mills theory is a remarkable relation
between string theory and quantum field theory, and it is important to extend it
to situations of less supersymmetry.  As with all dualities, information
flows in both directions.  In one direction, the less supersymmetric
gauge/gravity duals give quantitative solutions to strongly coupled gauge
theories with confinement and chiral symmetry breaking.  In the other, they
give information about string compactifications with realistic amounts of
supersymmetry as well as large warp factors.

In this talk I will focus on $D=4$, ${\cal N} = 2$ duals.  Systems with ${\cal
N} = 2$ supersymmetry are often studied as a step towards ${\cal N} = 1$,
taking advantage of the extra supersymmetry.  Happily, recent progress in ${\cal
N} = 1$ gauge/gravity duals has been quite rapid, so this motivation is no
longer as great.  Nevertheless, it is interesting to study the ${\cal N} = 2$
systems, both for the perspective that they give on ${\cal N} = 1$ and because
there are some interesting open questions.

The supersymmetry and conformal
invariance can be broken in various ways.  I will focus on breaking by the
combination of orbifolding and fractional brane fluxes, but will
also discuss briefly breaking by small perturbations (the latter is based on
work with Buchel and Peet\cite{BPP}).
A central issue is that the breaking of the conformal invariance of the AdS
space often leads to the horizon being replaced by a naked singularity.  It has
now been seen in many examples\cite{enhan,PS,KS,maldanun} that the singularity
is not actually present, though the mechanism that eliminates it varies from
case to case.  There is a particular phenomenon, the enhan\c
con, that is characteristic of ${\cal N} = 2$ systems.\cite{enhan}

In section~2 I review singularity resolution by the enhanc\c on.  In section~3
I review gauge/gravity duality with orbifolding and fractional branes, and
describe an ${\cal N} = 2$ analog of Seiberg duality.  In section~4 I study
the
${\cal N}=2$ version of the Klebanov-Strassler solution,\cite{KS} showing that
there is a duality cascade in the UV as in the ${\cal N}=1$ case, but that the
infrared behavior  is the characteristic ${\cal N}=2$ enhan\c con.  
Finally, I discuss dualities relating various solutions, and some open
issues.\footnote{I should point out that Igor Klebanov, with a series of
collaborators, has pioneered the study of this fascinating system.  Much of
my talk is a review of his work, with a few new ingredients --- the exact
${\cal N} = 2$ solution, the analysis of the ${\cal N} = 2$ duality
cascade and singularity resolution, and some of the final remarks.  See also
the talks by Klebanov and Strassler.

After this was written I learned of the recent ref.~7, with which it has 
substantial overlap.  Ref.~8, which has since appeared, presents a new ${\cal
N} = 2$ supergravity dual; how it is related to the solutions described in
section~5.2 is not yet clear.
}

\textheight=7.8truein
\setcounter{footnote}{0}
\renewcommand{\thefootnote}{\alph{footnote}}

\section{The Enchan\c con}
\noindent
For convenience let us focus on pure ${\cal N}=2$ $SU(N)$ Yang-Mills theory.
The Coulomb branch is $N$ complex dimensional.  Correspondingly the brane
system is made up of $N$ constituents, each moving in a two-dimensional
transverse space: these are 3-branes in six dimensions, with four dimensions
having been compactified or reduced.

When the constituents are well-separated, the metric $g_{ij}$ on the moduli
space of their positions is flat, but at sufficiently small separations
it becomes negative.  This is unphysical: the condition
\begin{equation}
g_{ij} > 0
\end{equation}
thus excludes a region in the interior of the moduli space.  On the boundary
of this region the metric has a zero, so at least one of the 3-branes has
vanishing tension.  A 3-brane of vanishing tension is of course a rather
special object, and so it must be at a spacetime point of nontrivial infrared
physics. In the $2+1$ dimensional case it is a point of enhanced gauge symmetry,
hence the name {\it enhan\c con}\cite{enhan} for such point; in the $3+1$
dimensional case there are tensionless strings.  We will also use the term
enhan\c con for the codimension one surface in moduli space where the metric
has a zero.

Generically the metric has only one zero, but there are points in the
$N$-dimensional moduli space where multiple 3-branes are tensionless.  If we
try to bring the 3-branes as close together as possible, they will all be
tensionless, and their positions will map out some closed curve in spacetime
--- the enhan\c con points fill out, in the large-$N$ limit, a closed curve.

In refs.~3 and 9 the supersymmetry was broken to ${\cal N}=2$ by different
means, but the IR physics in both is governed by the enhan\c con.
In ref.~3 the constituents were D7-branes wrapped on K3, with ${\cal
N}=4$ broken to ${\cal N}=2$ by curvature.  For simplicity it was assumed that
these were distributed on a ring; as its radius was reduced
they all became tensionless simultaneously.  In ref.~9 the breaking to 
${\cal N}=2$ was by a perturbation of the $AdS_5 \times S^5$ background,
corresponding\cite{WGKP} to a mass perturbation of the ${\cal N}=4$ theory.
A family of solutions were obtained by the lift of five-dimensional
solutions, correspond to a one-parameter ($\gamma < 0$) curve in moduli space. 
At the limiting value $\gamma = 0$ the constituent D3-branes all lay on
enhan\c on points, which in this case formed a segment of the real
axis.\cite{BPP}

The enhan\c con has a simple gauge theory interpretation.\cite{enhan,BPP,EJP}
In the ${\cal N} = 2$ theory the metric on moduli space receives corrections
only at tree level, one loop, and nonperturbatively.  The whole is positive
definite, but the perturbative part alone goes negative in parts of
moduli space.\cite{SW}
The instanton corrections are of
order $e^{-O(g_{\rm YM}^{-2})}$.  Since $g_{\rm YM}^2 = O(N^{-1})$, the
perturbative part is accurate until one reaches a point on moduli space where 
$g_{\rm YM}^{-2} \to 0$, at which point the corrections must suddenly become
important. Noting that the metric on moduli space is essentially $g_{\rm
YM}^{-2}$, the enhan\c con surface in moduli space is precisely where the
instanton effects become important.  Of course, the moduli space does not end
there. By holomorphy it must continue into the interior, but as argued in
ref.~3 it can no longer be interepreted as the moduli space of the positions of
pointlike constituents: naive continuation of the supergravity solution into
the interior would be singular.  Rather, the moduli represent the internal
state of the theory at the enhan\c con; it would be interesting to make this
more precise.

In other words, the enhan\c con is the large-$N$ manifestation of
the Seiberg-Witten curve,\cite{SW} which takes the form\cite{SWN}
\begin{equation}
y^2 = x^{2N} - \phi^{2N} + \Lambda^{2N}
\end{equation}
for a ${\bf Z}_N$-invariant Higgs v.e.v.\ of magnitude $\phi$.  For $\phi >
\Lambda$, the $\Lambda^{2N}$ term is negligible and the physics is
perturbative; for $\phi < \Lambda$ the instanton correction $\Lambda^{2N}$
is dominant.  Note that for $\phi = \Lambda$ the curve degenerates and one
has an Argyres-Douglas point; it would be interesting to see whether this
has any special geometric interpretation in the enhan\c con
context.\footnote{This remark is due to Cumrun Vafa.}

\section{Orbifolds and Fractional Branes}

\subsection{D3-Branes on the ${\bf Z}_2$ orbifold}
\noindent
The symmetries of $AdS_5 \times S^5$ can be broken by a perturbation which
becomes linear at the boundary, corresponding to a relevant perturbation of
the gauge theory Hamiltonian,\cite{WGKP} or by a large deformation at the
boundary.  Generally the latter corresponds to placing the D3-branes at a
singular point of the transverse
space.\cite{kacsil,kehag,klebwit,morpless,ahamal}  I will consider the
simplest case, of a ${\bf Z}_2$ orbifold
\begin{equation}
R:\quad x^{4,5} \to x^{4,5} \  , \quad x^{6,7,8,9} \to - x^{6,7,8,9}\ .
\end{equation}
This breaks half of the supersymmetry, leaving ${\cal N} = 2$.  In this
section I will review the gauge/gravity duality for this
orbifold.\cite{kacsil,klebnek}

On the string side one restricts as usual to  
${\bf Z}_2$-invariant configurations and adds in twisted sectors, states
localized on the fixed plane $x^{6,7,8,9} = 0$.  On the gauge side, the
massless open string spectrum is obtained by standard
technology.\cite{gimpol,quiver}  To describe $N$ D3-branes on the orbifolded
space we also need their $N$ images (fig.\ 1a), giving a total of $2N$
Chan-Paton degrees of freedom labeled $i \in 1,\ldots,2N$.
\begin{figure}[htbp]
\vspace*{13pt}
\epsfbox{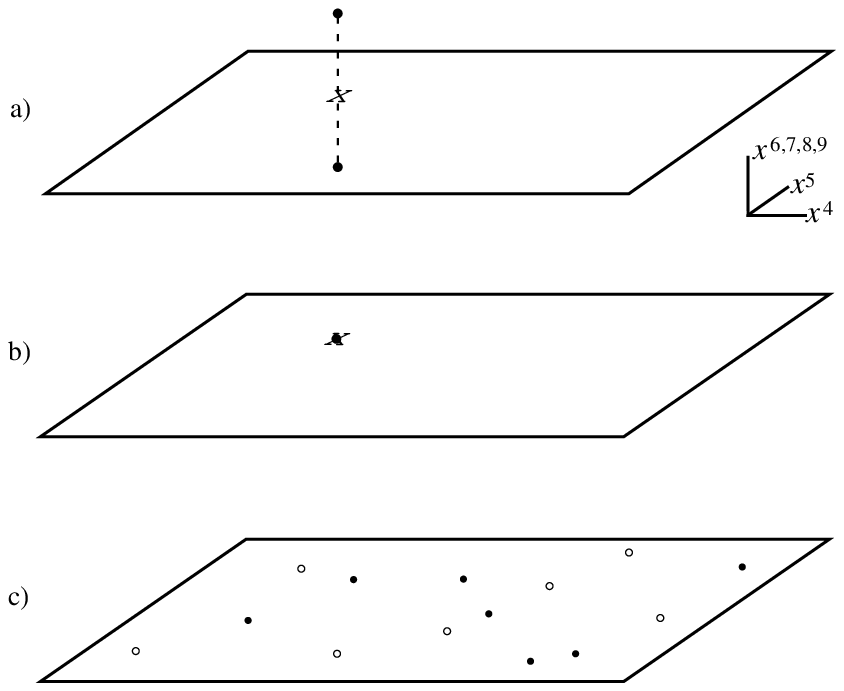}
\vspace*{13pt}
\fcaption{a) $N$ D3-branes above the fixed plane, and their images below.
b) All $2N$ branes+images coincident on the plane.  c) $N$ D5-branes (closed
circles) and $N$ anti-D5-branes (open circles) separated on the fixed plane.
}
\end{figure}
The orbifolding
acts both on the oscillator state $\psi$ of an open string and on its Chan-Paton
degrees of freedom:
\begin{equation}
R | i,j,\psi \rangle = \gamma_{ii'} \gamma_{jj'} | i',j',\hat R \psi \rangle\ ,
\end{equation}
where the matrix $\gamma$ interchanges each D3-brane with its image,
\begin{equation}
\gamma = \biggl[ \begin{array}{cc} 0 & I_N \\ I_N & 0 \end{array} \biggr]\ .
\label{basis1}
\end{equation}
The orbifold projection retains states even under $R$.

We are interested in the case that the D3-branes (and so their images) lie in
the fixed plane, as in fig.\ 1b.  To analyze this case it is useful to go to a
different basis for the Chan-Paton factors,
\begin{equation}
\gamma = \biggl[ \begin{array}{cc} I_N & 0 \\ 0 & - I_N \end{array} \biggr]\
.\label{basis2}
\end{equation}
In this basis, states in the diagonal blocks survive the projection if their
oscillator state is even under $\hat R$d and states in the off-diagonal blocks
survive if it is odd.  This leaves the massless states
\begin{equation}
\biggl[ \begin{array}{cc} A^{\mu},\, X^{4,5}  & X^{6,7,8,9} \\ X^{6,7,8,9} & 
 A^{\mu},\, X^{4,5} \end{array} \biggr] \to 
\biggl[ \begin{array}{cc} A^{\mu},\, \phi  & A_{1,2} \\ B_{1,2} &
\tilde A^{\mu},\, \tilde \phi \end{array} \biggr] \label{openstates}
\ .
\end{equation}
In the second form we have collected the scalars into complex pairs.  The
gauge group is $U(N) \times U(N)$, with an adjoint ${\cal N} = 2$ vector
multiplet and two $({\bf N},\OL{\bf N}) + (\OL{\bf N},{\bf N})$ hypermultipets.
The notation~(\ref{openstates}) corresponds to ${\cal N} = 1$ 
multiplets.

The overall $U(1)$ decouples from the bulk dynamics, so the orbifolded string
theory is the dual to the ${\cal N} =2$ $SU(N) \times SU(N) \times U(1)$ gauge
theory with two bifundamental hypermultiplets.  Not surpisingly, since there
is still an $AdS_5$ factor, this is a conformally invariant
theory.\cite{kacsil,LNV}$^{,}$\footnote
{To be precise, the $U(1)$ is not conformal but IR free.\cite{fuchs}  How this
is manifests on supergravity side is a puzzle, because of the $AdS_5$ factor.}

\subsection{Fractional branes}
\noindent
The system becomes more interesting when generalized as follows.  In the
basis~(\ref{basis2}), we can generalize to blocks of different size,
\begin{equation}
\gamma = \biggl[ \begin{array}{cc} I_{N+M} & 0 \\ 0 & - I_N \end{array}
\biggr]\ .\label{basis3}
\end{equation}
This corresponds to a $U(N+M) \times U(N)$ gauge theory with two bifundamental 
hypermultiplets.
This $\gamma$ can no longer be brought to the form~(\ref{basis1}).  A
$2N$-dimensional submatrix can be, and so $N$ D3-branes can be moved off the
fixed plane, but this leaves $M$ `half-D3-branes.'  The `half' is because these
do not have images, and so for example couple with half the strength of a
D3-brane to closed string states.

We can also form these half-D3-branes directly in the $U(N) \times U(N)$
system.  The vevs of $\phi$ and $\tilde \phi$ give $2N$ independent
coordinates in the fixed plane, corresponding to the $N$ D3-branes separating
into $2N$ half-D3-branes.  These are of two types (fig.\ 1c), according to
whether $\gamma$ acts as $+1$ or $-1$ on the given Chan-Paton index.

These half-branes have a simple interpretation.\cite{fractional}
There are five massless scalar twisted states in the IIB theory.  Three
correspond to blowups of the fixed point $x^{6,7,8,9} = 0$.
Hidden in the fixed point is a zero-size $S^2$, which
acquires a finite size when the fixed point is blown up into a smooth space.
The other two scalars are $\theta$-parameters from the NS-NS and R-R two-form
potentials integrated over the $S^2$,
\begin{equation}
\theta_{\rm B} = \frac{1}{2\pi\alpha'} \int_{S^2} B_{\it 2}\ ,\quad
\theta_{\rm C} = \frac{1}{2\pi\alpha'} \int_{S^2} C_{\it 2}\ .
\end{equation}
c
The two types of half-brane correspond to D5-branes or anti-D5-branes wrapped
on this $S^2$.  Being D-branes, these should have a (magnetic) coupling to
$\theta_{\rm C}$.  This coupling is given by a disk amplitude with a single
vertex operator.  The vertex operator has a branch cut extending to the
boundary, and includes a factor of $\gamma$ acting on the boundary state.
Thus the coupling is proportional to the trace of $\gamma$, and so is zero for
a full D3-brane, but $\pm 1$ for the two types of half-brane.

The twisted state coupling measures the D5 charge.  The half unit of D3-brane
charge can be understood as follows.  A D5-brane couples to the Ramond
background as 
\begin{equation}
\int ( C_{\it 6} + B_{\it 2} \wedge C_{\it 4} ) \ .
\end{equation}
The first term is its magnetic coupling to $C_{\it 2}$.  The second is a D3
charge equal to $\theta_{\rm B}/2\pi$, and the free orbifold CFT is
known\cite{aspinwall} to corresponds to $\theta_B = \pi$; indeed, this is
probably the most physical way to derive that fact.  Thus, fig.\ 1c
corresponds to $N$ D3-branes separating into $N$ [D5 + \half D3] plus
$N$ [$\OL{\rm D5}$ + \half D3].  The more general matrix $\gamma$ corresponds
to 
\begin{equation}
(N+M)\  \Bigl[ {\rm D5} + \half {\rm D3} \Bigr] \ \oplus\ N\ \Bigl[\OL{\rm D5}+
\half {\rm D3} \Bigr]
\label{constit}
\end{equation}
Thus, the $U(N+M)$ gauge factor is associated with the wrapped D5-branes, and
the $U(N)$ factor with the wrapped anti-D5-branes. 

\subsection{An ${\cal N}=2$ Seiberg duality}
\noindent
The final issue to understand is the effect of varying $\theta_{\rm B}$,
particularly on the couplings of the two gauge groups.  First let us stay in
the range $0 \leq \theta_{\rm B} \leq {2\pi}$.  The
system~(\ref{constit}) evolves to
\begin{equation}
(N+M)\  \Bigl[ {\rm D5} + \slaf\theta_{\rm B} {\rm D3} \Bigr] \ \oplus\ N\
\Bigl[\OL{\rm D5}+
(1-\slaf\theta_{\rm B}) {\rm D3} \Bigr]\ .
\label{constit2}
\end{equation}
For these objects, the tension is simply proportional to the magnitude of the
D3-brane charge,
\begin{equation}
\tau = \tau_3 |Q_3|\ ,\quad \tau_3 = \frac{1}{(2\pi)^3 \alpha'^2 g_{\rm s}}\
,
\label{BPS}
\end{equation}
and so is respectively $\slaf\theta_{\rm B} \tau_3$ and 
$(1-\slaf\theta_{\rm B}) \tau_3$ for the constituents~(\ref{constit2}).
The D5-brane tension makes no contribution because the sphere on which the
D5-brane is wrapped has zero size.  Note that --- as long as all constituents
have the same sign of $Q_3$ --- the full system satisfies the BPS
property~(\ref{BPS}) and so is a BPS state, in spite of having both D5-branes
and anti-D5-branes. Of course, if the $S^2$ is blown up to finite size, the
BPS mass will have term involving $|Q_5|$ and the fully separated system of
fig.\ 1c will no longer be supersymmetric --- the fractional branes of
opposite $Q_5$ will bind into full D3-branes.

The gauge field action comes from expanding out the Born-Infeld Lagrangian
\begin{equation}
- \tau \sqrt{ - \det( G_{\mu\nu} + 2\pi\alpha' F_{\mu\nu} ) }\ .
\end{equation}
It follows that the gauge coupling is just proportional to the tension, and so
for the constituents~(\ref{constit2})\cite{LNV}
\begin{equation}
\frac{4\pi}{g^2_{SU(N+M)}} = \frac{\theta_{\rm B}}{2\pi g_{\rm s}}\ ,\quad
\frac{4\pi}{g^2_{SU(N)}} = \frac{2\pi - \theta_{\rm B}}{2\pi g_{\rm s}}\ .
\label{gauge}
\end{equation}

The angle $\theta_{\rm B}$ is a periodic variable, but the system that we are
studying undergoes an interesting spectral flow.  Consider increasing
$\theta_{\rm B}$ past $2\pi$.  The D3 charge of the anti-D-branes becomes
negative, so the system~(\ref{constit2}) is no longer BPS.  However, it can
remain BPS by rearranging into new contituents as $\theta_{\rm B}$ goes through
$2\pi$,
\begin{equation}
(N+2M)\  \Bigl[ {\rm D5} + (\slaf\theta_{\rm B} - 1) {\rm D3} \Bigr] \ \oplus\
(N+M)\
\Bigl[\OL{\rm D5}+
(2-\slaf\theta_{\rm B}) {\rm D3} \Bigr]\ .
\label{constit3}
\end{equation}
Thus the gauge group changes as well, from $U(N+M) \times U(N)$ to $U(N+2M)
\times U(N+M)$.  This is the same Seiberg duality\cite{seiberg} that appears
in ${\cal N} = 1$.\cite{KS}  Note that we are treating $\theta_{\rm B}$ as a
constant, but the gauge couplings will run with energy; we will take
account of this in the next section.

For $M=0$ there is no spectral flow and $\theta_{\rm B}$ is simply a periodic
variable; it is remarkable that the inverse gauge coupling is thus also a
periodic variable,\cite{period} which can be smoothly taken to 0 and then back
to positive values.  For $M \neq 0$ it would be interesting to understand
better the nature of the flow.  Note that there is a $U(N+M)$ factor on both
sides, whose gauge coupling is finite and continuous
through the transition.
Note also that we can think of $U(N+2M)$ as related to $U(N)$
through Higgsing by the scalar field $\phi$ in the vector multiplet,
with the $U(N+M)$ group as a spectator.\footnote{The $U(N+2M) \times
U(N+M)$ can also be broken to $U(N+M) \times U(N) \times U(1)$'s
by a combination of $\phi$ and $\tilde\phi$, or by the hypermultiplets $A_i$ and
$B_i$.  The former most closely reflects the actual distribution of D-brane
charges, but it is unlikely that the rearrangement between
constituents~(\ref{constit2}) and~(\ref{constit3}) can be understood in terms of
classical Higgsing. I would like to thank Ofer
Aharony for extensive discussions of this point.}$\ $ However, this simple
picture is probably not correct: note that the $U(N+M)$ factor initially acts on
the D5-brane Chan-Paton factors, but after the transition it acts on the
anti-D5-brane Chan-Paton factors; simple Higgsing would not have this effect.
Also, we might expect some form of electric-magnetic duality to relate the two
sides;\cite{seiberg} it is not clear how to see this.

\section{The Supergravity Dual}
\noindent
We now consider the supergravity solution corresponding to a distribution of
D5- and D3-branes.  We first consider the case that the D5-branes can be
treated as a perturbation, as in ref.~19.  With the three-form flux having
components only transverse components, its equation of motion takes the simple
form
\begin{equation}
d [ Z^{-1} ( *_6 G_{\it 3} - i  G_{\it 3})] = 0\ ,
\end{equation}
where $Z$ is the harmonic (with sources) function appearing in the D3-brane
background and
$G_{\it 3} = F_{\it 3} - \tau H_{\it 3}$.  In the present case, $G_{\it 3} =
dA_{\it 2}$ with
\begin{equation}
A_{\it 2} = \theta(x^4,x^5) \omega_{\it 2}\ ,
\end{equation}
where $\theta = \theta_{\rm C} - \tau\theta_{\rm B}$ and $\omega_{\it 2}$ is
the two-form localized at the fixed point, with unit integral over the small
two-sphere.  In the orbifold limit $\omega_{\it 2}$ has delta function
support, but expressions can be regularized by slightly blowing up the fixed
point.  The equation of motion then takes the form
\begin{equation}
\partial_z ( Z^{-1} \partial_{\bar z} \theta ) = 0
\end{equation}
in terms of $z = x^4 + i x^5$.

A simple class of solutions is that $\theta$ be analytic, and this moreover
is the condition for ${\cal N} = 2$ supersymmetry.  The wrapped D5-branes
are magnetic sources for $\theta_{\rm C}$, so\footnote
{I would like to thank the authors of ref.~7 for correcting a factor of 2 in
this and the subsequent equations.}
\begin{equation}
\theta = - 2i \sum_i q_i \ln (z - z_i)\ ,
\end{equation}
where $q_i = \pm 1$ and
$z_i$ are the D5 charge and position of each wrapped 5-brane. The branch cut in
the logarithm corresponds to the equivalence
$\theta_{\rm C} \cong \theta_{\rm C} + 2\pi$, with a factor of 2 arising because
this is the self-intersection number of the $S^2$.  This is all as in ref.~19. 
The one new ingredient I have to add is the exact solution with backreaction.
With $\theta$ analytic, the solution is in the class found in ref.~28: the
three-form flux acts precisely as a D3-brane source in the other field
equations, so there backreaction appears in a modified harmonic function
$Z'$.  That is, in addition to a source from any explicit D3-branes, there is
a  source on the fixed plane, proportional to $|\partial_z \theta|^2$.
For example, the source for $dF_{\it 5}$ is proportional to $H_{\it 3}
\wedge F_{\it 3}$.

Consider now $M$ D5-branes at the origin, so that
\begin{equation}
\theta =  -2i M \ln z\ .
\end{equation}
The first point to notice is that $\theta_{\rm B}$ is a function of radius,
$gM \ln r$.  Through eq.~(\ref{gauge}) this translates into a running of the
couplings, which agrees with the ${\cal N} = 2$ beta function.\cite{klebnek}
Thus there is a duality cascade, with the successive
duality steps separated in scale by a factor $e^{2\pi/gM}$; note that the
exponent is small in the supergravity regime.

Now consider the backreaction.  The effective D3 charge density is
proportional to $|\partial_z \theta|^2 = M^2 |z|^{-2}$ and so logarithmically
divergent both in the UV and IR.  The normalization is such that the total
D3 charge between two duality cascades is exactly $M$.  This is as expected:
the size of the gauge group matches the D3 charge, but as in the
Klebanov-Strassler (KS) solution\cite{KS} these D3-branes are realized as
flux, not mobile D-objects (more on this in the next section).

The UV divergence is as in the KS solution, reflecting the unbounded growth
of the gauge group.  One difference is that in the ${\cal N}=1$ KS case,
there is no known way to terminate the cascade in a four-dimensional field
theory, while in the ${\cal N} = 2$ case it is possible: the $U(N+M) \times
U(N)$ theory can be obtained as the low energy limit of a spontaneously
broken $U(N+M) \times U(N+M)$ theory, which is conformal in the UV.  In
terms of branes, beginning with $N+M$ D3-branes we can pull $M$
anti-D5-branes out to long distance.

The IR divergence, however, is a problem.  It can be
`renormalized,' leaving a finite backreaction, by adding a negative
infinite D3 charge at the origin.  However, this solution is singular.  For
a sufficiently small $r$ the total D3 charge within is negative, and so the
warp factor $Z'$ changes sign.  Where it goes to zero, the metric
coefficients $g_{\mu\nu}$ vanish as well.  This is the characteristic
`repulson' singularity,\cite{repul} whose resolution is the enhan\c
con:\cite{enhan} the configuration with all the D5-branes at the origin is
unphysical.  

Let us see this as in ref.~3, by starting with the constituent
branes dispersed and trying to contract the distribution.  For convenience
we will regulate the UV also, as described above: we start with a ring of
$N+M$ D3-branes at radius $r_0$, in the orbifold limit $\theta_{\rm B} =
\half$.  Leave $M$ anti-D5-branes at $r_0$ and contract the rest.  When the
inner ring is at radius $r'$ we have 
\begin{equation}
\theta_{\rm B} = \left\{
\begin{array}{ll} \half\ , \quad& r > r_0 \\[2pt]
\half - 2gM \ln (r_0/r)\ , \quad& r_0 > r > r' \\[2pt]
\half - 2gM \ln (r_0/r')\ , \quad& r' > r 
\end{array} \right.
\ .
\end{equation}
The inner ring is initially composed of $N+M$ D5-branes and $N$
anti-D5-branes, but as it moves inward and $\theta_{\rm B}(r')$ passes
through integer values, the constituents reorganize as described in
the previous section and the system cascades downwards, the number of branes
of each type decreasing by $M$ at each step. 

For $N$ a multiple of $M$, after $K = N/M$ steps there remain $M$ D5-branes.
When these reach the enhan\c con radius
\begin{equation}
r_{\rm e} = r_0 \exp[ - \pi (K + \half)/ gM]
\end{equation}
they become massless and the contraction must stop.  At this point, all of
the D3-brane charge of the inner ring has been stored in the three-form flux
background.  The density is nonnegative everywhere and there is no repulson
singularity.  If $N$ is not a multiple of $M$, there remain at the end some
whole D3-branes.

In summary, the ${\cal N} = 2$ fractional brane system has several features in
common with the ${\cal N} = 1$ system, but the singularity resolution and IR
physics is the same as in the ${\cal N} = 2$ system considered in ref.~3. 
In the conclusions we will discuss some relations among these systems.

\section{Open Questions}

\subsection{Exotic Physics}
\noindent
An important open question in this and related systems is the nature of the
low energy physics, the physics of the enhan\c con ring.\footnote
{See also the talk by Clifford Johnson.}\ \ \ 
The final
D5-branes sit at a radius where $\theta_{\rm B}$ is an integer and so there
are tensionless strings.  Thus the effective low energy theory may be rather
exotic.  Note that in the present case there are additional low energy
degrees of freedom on every ring where $\theta_{\rm B}$ passes through an
integer.  As the constituents move inward and pass through a radius of
integer $\theta_{\rm B}$, they rearrange,
for example from the form~(\ref{constit3}) to~(\ref{constit2}), and the
number of moduli drops by $2M$.  The dimension of the ${\cal N} = 2$ 
moduli space cannot change, so these moduli must remain as degrees of
freedom of the low energy theory on the ring.  Thus the system exhibits an
interesting analog of `spin-charge separation': the D3-brane charge is
spread throughout space in the form $F_{\it 3} \wedge H_{\it 3}$, while the
D3 moduli are localized at discrete radii.

A related question concerns the dual at very large $g$.  In the ${\cal N} =
4$ system, the effective description at $g \ll 1/N$ is a perturbative field
theory, and at $g \gg N$ it is the dual field theory; in between is the
supergravity description.  In the present case, the description at $g \gg N$
is the field theory on wrapped NS5-branes, and it is not clear what this
should be --- there seems to be no simple orbifold interpretation.  Note that
in this limit the rings of integer
$\theta_{\rm B}$ are very closely spaced and so there are many exotic low
energy degrees of freedom.

\subsection{Related Solutions}
\noindent
It is interesting to take the $T$-dual on an `angular' $U(1)$ at the ${\bf
Z}_2$ singularity.\cite{tdual,enhan}  This gives a pair of NS5-branes,
extended in the 12345-directions and separated in the periodic 6-direction,
with $M$ D4-branes stretched between.  The NS5-branes bend logarithmically
away from one another and repeatedly intersect due to the periodicity ---
these intersections are at the radii of exotic low energy physics.  

Now
imagine the same brane configuration, rotated so that the separation and
bending are in the 7-direction, while the 6-direction remains periodic.  
The shape of the branes is the same, though they no longer intersect.
This 
$T$-dualizes back to a different solution: the 6-separation is $T$-dual to
$\theta_{\rm B}$, while the 7-separation is $T$-dual to the volume of the
$S^2$. Thus one obtains a solution in which the $S^2$ does not remain at zero
volume but expands logarithmically.  This is the original context in which
the enhancon was discussed,\cite{enhan} though the focus there was a
yet-different $T$-dual form in which the volume of an entire K3 varies
logarithmically.

Rotating one of the NS5-branes into the 12389 directions gives a
$T$-dual to the ${\cal N} = 1$ KS solution.\cite{conit,KS}  Thus the ${\cal N}
= 1$ and~2 singularity resolutions can be
continuously connected, even though one
involves supergravity fields only and the other involves branes.  Further, if this entire brane configuration is now
rotated from the 6- into the 7-direction, one obtains a solution with vanishing
NS-NS flux but a logarithmically growing $S^3$ --- presumably the
Chamseddine-Volkov solution,\cite{CV} interpreted by Maldacena and Nu\~nez as dual to a confining gauge theory.\cite{maldanun}  The CV-MN and KS duals are
different in the UV, the former being six-dimensional and the latter
four-dimensional but with an unbounded gauge group; evidently the same number of
degrees of freedom can be arranged in these two different ways.

The solutions in the 6-direction have constant dilaton and
imaginary-self-dual three-form flux; the solutions in the 7-direction have
real three-form flux with a related dilaton gradient.  By rotating to an
intermediate angle and taking the $T$-dual, one will obtain a solution of
more general form.

\nonumsection{Acknowledgements}
\noindent
I would like to thank Ofer Aharony, Alex Buchel, Ehud Fuchs, Mariana Grana, Steve
Gubser, Clifford Johnson, Andreas Karch, Igor Klebanov, Nikita Nekrasov, Amanda
Peet, Matt Strassler, and Cumrun Vafa for discussions and collaboration.
This work was supported by National Science Foundation grants PHY99-07949 and
PHY97-22022.

\nonumsection{References}

\end{document}